\documentclass[aps,twocolumn,superscriptaddress,showpacs]{revtex4}
\usepackage{graphicx}

\makeatletter
  \def\@oddfoot{{\sl \rightmark}\hfil Abrams, Lee, Obukhov, p.\ \rm\thepage}
\makeatother

\begin{document}

\title{Collapse Dynamics of a Polymer Chain: Theory and Simulation}

\author{Cameron F. Abrams}
\affiliation{Max-Planck-Institute for Polymer Research, P.O. Box 3148,
  D-55021 Mainz, Germany}
\author{Namkyung Lee}
\affiliation{Max-Planck-Institute for Polymer Research, P.O. Box 3148,
  D-55021 Mainz, Germany}
\author{Sergei Obukhov}
\affiliation{Department of Physics, University of Florida,
  PO-Box 118440, Gainesville, Florida, 32611-8440} 
\date{16 January 2002}

\begin{abstract}
  We present a scaling theory describing the collapse of a homopolymer
  chain in poor solvent.  At time $t$ after the beginning of the
  collapse, the original Gaussian chain of length $N$ is streamlined
  to form $N/g$ segments of length $R(t)$, each containing $g \sim t$
  monomers.  These segments are statistical quantities representing
  cylinders of length $R \sim t^{1/2}$ and diameter $d \sim t^{1/4}$,
  but structured out of stretched arrays of spherical globules.  This
  prescription incorporates the capillary instability.  We compare the
  time-dependent structure factor derived for our theory with that
  obtained from ultra-large-scale molecular dynamics simulation with
  explicit solvent.  This is the first time such a detailed comparison
  of theoretical and simulation predictions of collapsing chain
  structure has been attempted. The favorable agreement between the
  theoretical and computed structure factors supports the picture
  of the coarse-graining process during polymer collapse.

\end{abstract}

\maketitle

The collapse transition of a single polymer molecule is for several
reasons an interesting physical problem.\cite{degennes85,grosberg93,timoshenko95b+kuznetsov95c,buguin96,byrne95,pitard98+99,chang01,milchev94,chu95}
Understanding the collapse transition is foremost a precursor to
understanding protein folding,\cite{pande97} and how other
biomolecules react to changes in environment.  Furthermore, a better
physical picture of the collapse transition will enhance the ability
of nanoscale micromanipulation techniques to provide a way to
experimentally approach many such problems.\cite{chu95,hagen96}

The earliest consistent theoretical picture of the collapse transition
stems from de Gennes.\cite{degennes85} In this scenario, when the
temperature is shifted by $\Delta T$ from $\theta$ conditions, a
Gaussian coil starts to aggregate, resulting in a uniformly dense
sausage-like shape.  As time goes on, the minimization of interfacial
area drives this sausage-like shape to thicken and shorten until, at
the final stage, a globule is formed.  More recent work has shown that
such uniform sausage-like shapes are highly unstable in solvents due
to the capillary instability, which selects for ``pearl-necklace''
structures over uniform sausages.\cite{byrne95} Attempts made to
incorporate the capillary instability in the description of the
collapse of a polymer chain,\cite{halperin00,klushin98} however, have
so far resulted in no universally accepted theoretical description,
particularly in prediction of how the time for collapse scales with
chain length.  The purpose of this paper is to present a theory based
on a new idea for the mechanism of the collapse transition, and to
support this theory by comparing directly to results of large-scale
molecular dynamics simulations.

\begin{figure}
\includegraphics[width=7cm]{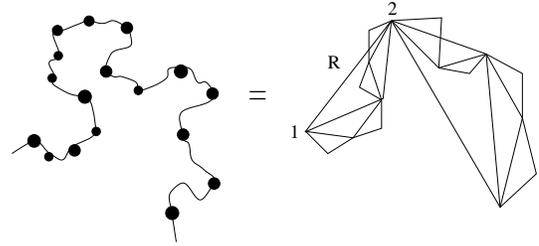}
\vspace{5mm}
\caption{Equivalence between a Gaussian coil and a Gaussian fractal.}
\label{fig:gauss}
\end{figure}

Consider a Gaussian coil representing a polymer in $\theta$ solvent.
We may think of this coil as a Gaussian fractal, created by a
heirarchical scheme, demonstrated in the right side of
Fig.~\ref{fig:gauss}.
We subject this system to an instantaneous drop in temperature, which
quenches the system rapidly into poor solvent conditions.  This
effectively introduces monomer-monomer attraction.  We assume this
quench is so deep that the initial nucleation of the first globules of
the collapsed polymer phase appear instantaneously.  (Otherwise,
addition time scales and regimes would enter the
problem.\cite{buguin96,klushin98,privatecomm}) The system evolves
dynamically according to microscopic equations of motion of chain
segments.  The segments are straightened arrays of globules, and
further straightening streamlines them at larger length scales. This
leads to the overall coarse-graining of the chain, on a length scale
that is growing with time.  This process is the reverse of the
heirarchical Gaussian fractal generation.  

\begin{figure}
\includegraphics[width=3in]{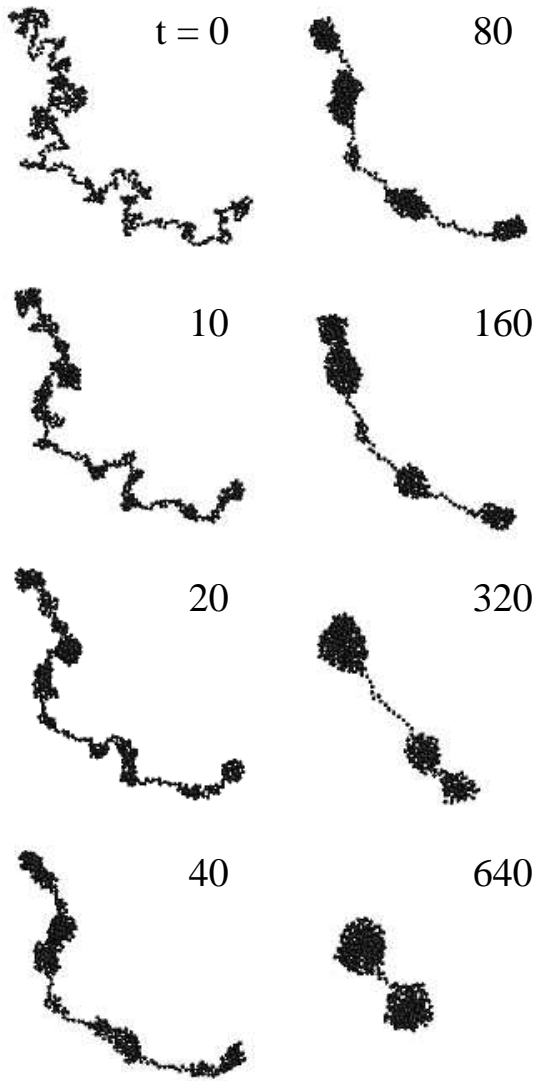}
\caption{Simulation snapshots of a collapsing $N$ = 512
  chain. Time $t$ is measured from the instant at which the solvent
  quality is changed from ``good'' to ``poor''.}
\label{fig:snaps}
\end{figure}

Because we view the original Gaussian chain made of $N$ monomers of
size $b_0$, we can define a length scale $R$ as the resolution at
which the chain appears as a Gaussian chain made of $N/g$
coarse-grained monomers, or segments, of size $b = b_o g^{1/2}$,
containing $g$ original monomers.  Viewing the collapsing chain at a
length scale smaller than $R$ reveals straightened segments.  However,
the overall large scale structure still remains a random walk of
segments.  This picture is clearly captured in snapshots from
simulations, which we discuss below; see Fig.~\ref{fig:snaps}.

If we neglect numerical factors associated with detailed shapes of the
objects being pulled through the viscous liquid, the hydrodynamic
equations of motion for coarse-grained monomers at scale $R$ can be
written:
\begin{equation}
  v = f / \eta_o A(R)
\label{eq:drag-velocity}
\end{equation}
where $\eta_o$ is solvent viscosity, $f$ is the chain tension, and $v$
is the drag velocity.  The quantity $A(R)$ measures the friction
experienced by objects due to the solvent.  For Zimm dynamics, $A(R)$
depends on the largest geometric dimension of the object: $A(R) \sim
R$. In Rouse dynamics, $A(R) \sim g$ regardless of its geometric size.
For now, we assume Zimm dynamics holds, and hence, the velocity scales
as $v \sim f/\eta_o R$.

We account for the chain tension $f$ as follows.  By bringing $g_l\sim
L$ monomers forming a string of length $L\leq R$ into a globule, one
gains free energy $\Delta E/k_BT = g_l\Delta T/\theta$.  Thus,
\begin{equation}
f  = \frac{\Delta E}{L}= - \frac{k_BT}{b_o}  \Delta T/\theta.
\end{equation} 
Note this force $f$ is independent of scale $R$.  In this estimation
of force, we neglect the surface term correction, $\sim
k_BT/b_og^{-1/3} (\Delta T/\theta)^{4/3}$, as it decays relatively
rapidly with increasing $g$.  There are also external forces acting on
each segment from the remainder of the chain.  But because the
segments form a random walk, the vectoral average of these external
forces acting on the ends of any segment is zero.  Hence, the
initially given Gaussian structure remains through the
coarse-graining, as the length of the segments increases.

The velocity $v$ of motion on each scale $R$ is controlled by
Eq.~\ref{eq:drag-velocity}. After a time $t \sim R/v \sim R^2$, in
which all segments move a distance $R$, the configuration on scale $R$
is ``relaxed.''  We can assume that all $g$ monomers of a subchain
collapse to a cylindrical configuration, or segment, of length $R(t)$
and diameter $d(t)$.  The volume of the cylinder $Rd^2$ should be
approximately equal to the total volume of $g$ monomers: $V \approx g
b_o^3$.  Since the largest dimension $R$ of the cylindrical object
remains the same as for initial Gaussian chain $R \sim g^{1/2}$, we
conclude that $g
\sim t$, and $d \sim g^{1/4}$.  Hence, we arrive at scalings $R \sim
t^{1/2}$ and $d \sim t^{1/4}$. (For Rouse dynamics, one obtains $R
\sim t^{1/3}$ and $d \sim t^{1/6}$).

It is important to realize that these cylindrical configurations exist
only in a statistical sense. Any uniformly dense ``cylindrical''
configuration would be immediately broken into an array of globules,
connected by strings, due to the capillary instability.\cite{siggia79}
The relaxation time for the capillary effect $t\sim d$, including the
coalescence of globules, is short, compared to streamlining time
$t\sim R^2$ (or $t\sim d^4$).  Thus, the formation of the
pearl-necklace structure (within the cylinder) does not require
modification of the streamlining description. Our cylinders are arrays
of globules characterized by an average diameter of $d$ and
inter-globule distance $d$, and overall array length $R$.  Mass
conservation then requires that each globule consists of $g^{3/4}$
monomers, and the total number of globules is $g^{1/4}$.

Furthermore, there is no energy barrier associated in bringing two
separated globules together. This is in contrast to models of other
authors, who maintain that globule interaction is the main mechanism
of relaxation.\cite{halperin00,klushin98} Instead of considering the
processes of coalescence of globules, and the competition in growth
and evaporation in detail, here we adopt the simple viewpoint that the
process of streamlining on scale $R$ results in the formation of
statistical cylinders of size $R$.  The contraction of each segment
drags the globules to the larger-scale cylinder.  The overall
structure is a random walk of such cylinders, which still displays a
Gaussian structure on larger scales until the end of coarse-graining
stage.

This mechanism allows direct derivation the time-dependent structure
factor $S(k,t)$ for a collapsing chain.  $S(k,t)$ is a useful
function, because it can be both measured experimentally and directly
computed from simulations. Below we report the structure factor $S(k)$
of a polymer chain made of cylinders, which in turn are made of linear
necklaces of spheres.  The structure factor can be written as a
product of structure factors of its elements:
\begin{equation}
\label{eq:sth}
S(k,t)=S_{gauss}[R(t)] S_{cyl}[R(t)] S_{glb}[d(t)].
\end{equation}
$S_{gauss}(k,t)$ is the structure factor of Gaussian chain consisting
of $N_1(t)=N/g(t)$ unit cylinder elements of average length $R(t)$.
Therefore,
\begin{equation}
S_{gauss}\left(R\right) = \frac{1}{N_1}\left\{\frac{2/N_1 \left[e^{-\alpha (N_1
+1)}-e^{-\alpha}\right]-e^{-2 \alpha}+1}
{(1-e^{-\alpha})^{2}}\right\}
\end{equation}
where $\alpha = \left[kR(t)\right]^2/3$.  For large $k$, we note that
this expression has a log-log slope of -2, as expected.
$S_{cyl}$ and $S_{glb}$ respectively are the structure factors of
cylinder with length $R(t)$ and a globule of size $d(t)$:
\begin{eqnarray}
\label{eq:scyl}
S_{cyl}(R,k)&=&\frac{2}{(kR)^2}\left[\cos kR + kR{\rm 
  Si}(kR)-1\right]\\
\label{eq:sglob}
S_{glb}(d,k)&=&\left[\frac{3}{(kd)^3}
\left(\sin kd - kd\cos kd \right)\right]^2
\end{eqnarray}
Si$(x)$ is the sine integral of $x$.  The length of cylinders $R(t)$
and the diameter of globules $d(t)$ are the parameters which follow
the coarse-graining rule suggested above. The large-$k$ log-log slopes
of the relations in Eq.~\ref{eq:scyl} and Eq.~\ref{eq:sglob} are -1
and -4 respectively.

As a test of our theory, we computed the structure factors for a
simulated polymer chain undergoing collapse.  The collapse of a
homopolymer chain in a poor solvent was simulated via molecular
dynamics (MD).  The prototype system is a single linear freely-jointed
polymer chain of length $N$ = 512 monomers, immersed in a solvent of
8$\times10^4$ simple particles, contained in a cubic box.  The total
particle number density, $\rho$, is 0.85 $\sigma^{-3}$, giving a box
dimension of 45.6 $\sigma$.  All particles interact via standard
Lennard-Jones potentials.  For ``good'' solvent conditions, the
solvent-solvent, monomer-monomer, and solvent-monomer pair
interaction, designated $U_g(r_{ij})$, is a shifted repulsive
Lennard-Jones potential, cut off at $r$ = 2$^{1/6}$ $\sigma$.  For
``poor'' solvent conditions, the solvent-solvent and monomer-monomer
pair interaction is the full Lennard-Jones potential,
$U_{LJ}(r_{ij})$, with a well depth $\epsilon$ of 1.0 $k_BT$, cut off
at $r$ = $r_c$ = 2.5 $\sigma$, while the solvent-monomer pair
interaction is kept as the repulsive $U_g$.  The shift $s$ of $U_{LJ}$
is set such that $U_{LJ}(r_c) = 0$.  This simple prescription was used
previously in smaller-scale homopolymer collapse
simulations.~\cite{chang01} Particles connected to each other along
the backbone of a chain are constrained via a harmonic potential with
an average length of 0.97 $\sigma$ and a standard deviation of 0.05
$\sigma$.

The equations of motion are integrated using the velocity-Verlet
algorithm, with a time step $\Delta t$ of 0.001 $\tau$, where
$\tau=\sqrt{m\sigma^2/k_BT}$, and $m$ = 1. Energy is measured in units
of $k_BT$, which prescribed that the particle diameter, $d_0$, be
exactly $\sigma$, for both monomers and solvent particles.  Thus the
total excluded volume fraction in the system is about 0.45.  

The procedure for a single collapse trajectory is as follows.  The
initial configuration was generated by growing a random walk of length
$N$ = 512 and step length $l$ = 0.97 $\sigma$, biased such that 1--3
monomers automatically have no overlap.  Then the remaining 80~000
solvent particles are added at random locations in the box.  Particle
overlaps are then removed using a previously detailed warmup
technique,~\cite{Abrams2001b} which incrementally increases the the
excluded volume diameter of the particles from zero to its full value
over about 5000 MD time steps in ``good'' solvent conditions.  During
this phase, a Langevin-type thermostat with friction $\Gamma = 5.0
\tau^{-1}$ is employed.  The amount of time used for this warmup was
kept as low as possible in order that the initial configurations
remain as Gaussian as possible without significant swelling due to the 
actual presence good solvent conditions. 

Immediately following the warmup, designated for convenience as $t=0$,
the quench into poor solvent conditions is performed via
instantaneously switching the monomer-monomer and solvent-solvent pair
interactions from $U_g$ to $U_{LJ}$.  The MD integration is run for
several thousand $\tau$, until the chain has fully collapsed into a
single globule. During this phase, no thermostat is employed, and the
temperature was observed for all runs to fluctuate around 1.0 by less
than 0.1\%.  The collapse is monitored using the radius of gyration of
the chain as a function of time. During this integration,
configuration snapshots are stored every ten thousand time steps (10
$\tau$) for later analysis.  All simulations were run in parallel on
256 processors using domain decomposition.~\cite{Puetz1998}

A total of 40 independent collapses were simulated, each beginning
with a unique chain conformation.  The time-dependent structure factor,
$S(k,t)$, was computed at times $t$ = 0, 10, 20, 40, 80, and 160
$\tau$, by averaging the static structure factor over all chains at
each of those times.  We note that the standard deviations were about
20\% of the average, which is consistent with large variances in size
among the different initial Gaussian chains.

\begin{figure}
\includegraphics[width=8cm]{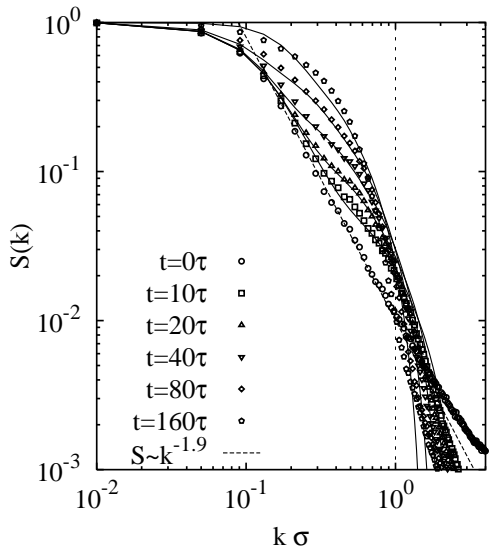}
\vspace{5mm}
\caption{The single chain time-dependent structure factor,
  $S(k,t)$. Comparison between theoretical predictions (lines) and
  simulation (points). The $t=0$ structure factor is measured over all
  chains immediately prior to the quench, and is best represented as
  $S\sim k^{-1.9}$, close to the theoretical $S\sim k^{-2}$.}
\label{fig:s(k)}
\end{figure}
The primary result of this work is a comparison of $S(k,t)$ predicted
by Eq.~\ref{eq:sth} and $S(k,t)$ computed from the MD simulations. In
order to make this comparison quantitative, we introduced the
necessary prefactors in the scaling relations $R \sim t^{1/2}$ and $d
\sim t^{1/4}$. We then performed a least-squares, conjugate gradient
minimization~\cite{Press-NR-ConjGrad} on the sum of squared errors
between the theoretical and computed $S(k,t)$ for all points for $k <
1$ and for all times simultaneously.  The resulting comparison is
shown in Fig.~\ref{fig:s(k)}.  Note that the theoretical and computed
$S(k,t)$ agree remarkably well.  In particular, both display the three 
asymptotic slopes of the structure factor components discussed
previously.  The theory and simulation are linked
through the following relations:
\begin{eqnarray}
R(t) & = & A t^{1/2}\\
d(t) & = & B t^{1/4}
\end{eqnarray}
with $A$ = 1.88(9) $\sigma/\tau^{1/2}$ and $B$ = 0.78(8)
$\sigma/\tau^{1/4}$.  The poorest agreement appears for the latest
time, perhaps signifying the end of the coarse-graining stage of the
collapse.  Due to the large inherent variability of the simulation
data, we do not report goodness-of-fit statistics.

To complete our picture, we discuss the late-stage collapse dynamics.
After the coarse graining stage, a chain becomes a single ``cylinder'',
where the globules are aligned linearly.  The terminal globules
experience net forces due to the finite size of the chain.  The late
stage is therefore characterized by ``end effects.''  The typical late
stage consists of two globules with connecting string between these
two.  In this regime, the dynamics are controlled by tension between
the two globules and the hydrodynamic friction experienced by the
pearls (not by the strings).  Both terminal globules move toward the
common center of mass by ``eating'' the string.  This procedure is
deterministic.  Considering the hydrodynamic friction for the sphere
with size $R_p$ is $\sim 6\pi\eta R_p$ and $R_p \sim N^{1/3}$, the
typical distance between the terminal globules should be $N^{1/2}$
(reminiscent of a Gaussian fractal), one can estimate the average
characteristic time for the late stage as $\tau_l\sim R/V \sim
N^{5/6}$ ($\tau_l^R \sim N^{3/2}$ for Rouse dynamics).  Future work
will offer a quantitative comparison of this picture with that
observed in ongoing MD simulations.

We have presented a new theory capable of describing the dynamics of
polymer collapse in poor solvent.  We have proposed that the initially
fractal chain recursively straightens on length scales $R(t)$, forming
an increasingly coarse-grained random walk of statistical cylinders of
length $R \sim t^{1/2}$ and diameter $d \sim t^{1/4}$, in accordance
with hydrodynamic equations of motion.  These cylinders are
statistical objects, whose underlying structure consists of
necklace-like configurations of randomly positioned globules connected
by strings.  This theory successfully incorporates the capillary
instability into a hydrodynamic framework. The theoretical time-dependent
structure factor is strongly supported by those computed from
large-scale MD simulations.  

\acknowledgments We thank T. Vilgis, V.  Rostiashvili, B. D\"unweg and
K. Kremer of the MPIP for thoughtful comments and discussions. We
thank the Rechnenzentrum of the Max-Planck-Gesellschaft in Garching,
Germany, for a generous allocation of Cray T3E CPU time.  M.  P\"utz
is gratefully acknowledged for his simulation code.  This work was
supported by the BMBF under grant No.  03 N 6015 (C.A.), and by a
grant from the DFG (N.L.).  One of us (S.O.)  is grateful for the warm
hospitality and support of the Max-Planck-Institut f\"ur
Polymerforschung (MPIP).

\end{document}